\documentclass[aps, 
floats,
preprint, 
showpacs,
nofootinbib,
tightenlines,
floatfix
]{revtex4}%
\usepackage{eurosym}
\usepackage{amsfonts}
\usepackage{amsmath}
\usepackage{graphicx}
\usepackage{subfigure}
\usepackage{relsize}
\usepackage{amssymb}%
{\catcode`\|=\active \gdef\Braket#1{\left<\mathcode`\|"8000\let|\bravert
{#1}\right>}}
\def\bravert{\egroup\,\vrule\,\bgroup}

\newcommand{\beq}{\begin{eqnarray}}
\newcommand{\eeq}{\end{eqnarray}}
\newcommand{\bw}{\begin{widetext}}
\newcommand{\ew}{\end{widetext}}

\begin{document}

\title{Extraction of $\alpha_s$ from deep inelastic scattering at large $x$}
\author{A. Courtoy}
\email{Aurore.Courtoy@ulg.ac.be}
\affiliation{
IFPA, AGO Department, Universit\'e de Li\`ege, B\^at. B5, Sart Tilman B-4000 Li\`ege, Belgium
}
\author{S. Liuti}
\email{sl4y@virginia.edu}
\affiliation{Department of Physics, University of Virginia, 382 McCormick Rd.,
Charlottesville, VA 22904, USA\\
}

\date{\today }

\begin{abstract}
We present an analysis of the role of the running coupling constant at the intersection of perturbative and nonperturbative QCD. 
Although the approaches that have been considered so far in these two regimes appear to be complementary to each other, a unified description might derive through the definition of the effective coupling, as they both provide ways of analyzing its freezing at low values of the scale.
We extract the effective coupling from all available experimental data on the unpolarized structure function of the proton, $F_2^p$,   at large values of Bjorken $x$, including the resonance region. We suggest that parton-hadron duality observed in this region can be explained if nonperturbative effects are included in the coupling constant.
The outcome of our  analysis is a smooth transition from perturbative to nonperturbative QCD physics, embodied in the running of the coupling constant at intermediate scales.
\end{abstract}

\maketitle

\noindent {\bf 1.}
Hard processes are described in QCD by envisaging a perturbative stage (PQCD) where a hard collision involving quark and gluons occurs, followed by a nonperturbative stage characterizing hadron structure.
For example, in  Deep Inelastic Scattering (DIS)  
the hard scattering part of the process, $\gamma^* q \rightarrow X$, 
can be presently described using splitting functions up to next-to-next-to-leading order (NNLO) \cite{Moch:2004pa,Vogt:2004mw}, and Wilson coefficients functions up to N$^3$LO
\cite{Vermaseren:2005qc}. The large distance part  is described by the Parton Distribution Functions (PDFs), 
which have to be extracted from fits of high energy experimental data. 
PDF parametrizations along with their uncertainties  have been obtained applying this framework up to NNLO, by a number of collaborations (see review in  \cite{Forte:2013wc}). 
NLO PDFs are also important for the comparison with data from, { \it e.g.}, collider processes which are only known up to 
NLO, and when using Monte Carlo event generators.
 
Although DIS stands out as the most accurately described process in a QCD factorized scenario, at large Bjorken $x$ large logarithms appear in the PQCD description that need to be resummed in order to ensure that the perturbative approach can be consistently extended. In this region power corrections of both kinematical (Target Mass Corrections, TMCs) and dynamical (Higher Twists, HT) origin also affect the extraction of PDFs. They are intertwined with Large $x$ Resummation (LxR), and the factorization framework is no longer expected to apply straightforwardly~\cite{Corcella:2005us}.
 
Early experimental observations indeed suggested that in specific kinematical regimes both the perturbative and nonperturbative stages arise almost ubiquitously, in the sense that the nonperturbative description tracks, in average, the behavior of the perturbative one. 
In inclusive $eP$ scattering at large $x$, a duality between the low-energy and high-energy behaviors of a same observable, {\it i.e.}, the unpolarized structure function, $F_2^p$ was  observed  by Bloom and Gilman who established a connection between the structure function in the nucleon resonance region and that in the deep inelastic continuum~\cite{Bloom:1970xb,Bloom:1971ye}. $F_2^p$, when averaged over  the resonance region, was found to be equivalent to the continuation of the deep inelastic curve in this region. This concept is known as {\it parton-hadron duality}: the resonances are not a separate entity but they are an intrinsic part of the scaling behavior of the structure function.
Note that here duality  implies an equality in average between resonances and scaling curve, the average being taken at fixed four-momentum transfer, $Q^2 \geq 1$ GeV$^2$,  over the same range in the scaling variable, $x$.  
This behavior can be taken as signaling a natural continuation of  the perturbative to the nonperturbative representation.

Although Bloom--Gilman duality was observed at the inception of QCD, quantitative analyses could be attempted only more recently, having at disposal the extensive, high precision data from Jefferson Lab~\cite{Melnitchouk:2005zr}. 

In this paper, to evaluate the impact of LxR,
we perform an analysis of inclusive $eP$ scattering data  at large $x$, including resonance data. 
To handle the latter, we follow closely the analysis of Ref.~\cite{Liuti:2011rw}, where the implications of  parton-hadron duality  were  explored
in the context of a systematic PQCD based analysis including besides LxR, TMCs and HT contributions or, more generally, the evidence for nonperturbative inserts, which are required to achieve a fully quantitative fit.
Here the relevant kinematical variables are: $x=Q^2/2M\nu$ ($M$ being the proton mass and $\nu$ 
the energy transfer in the lab system), the four-momentum transfer, $Q^2$, and the invariant mass for the proton, $P$, and virtual photon, $q$, system, $W^2=(P+q)^2$  
($W^2 = Q^2(1/x-1)+M^2$).  For
large values of Bjorken $x  \geq 0.5$, and $Q^2$ in the multi-GeV$^2$ region, one has 
$W^2 \leq 5$ GeV$^2$, {\it i.e.},  the cross section is dominated by resonance formation.

As first pointed out in Refs.~\cite{Liuti:2011rw,Bianchi:2003hi}, the extension of PQCD evolution to large $x$ must also include LxR effects. 
A consequence of  LxR is that as $x$ increases there is a shift to lower values of the scale at which $\alpha_s$ is calculated
 (see for instance  Ref.~\cite{Brodsky:1979gy} or the classical review in  Ref.\cite{Pennington:1982kr}).  
At large $x$ and low $W^2$, this shift requires a freezing of the coupling constant in the infrared region. 

The key result in the present work is that, by  turning this argument around, or by using the fact that our analysis, through LxR, is regulated by the value of the QCD coupling in the infrared region, we can extract such coupling from experimental data. 
In our analysis canonical higher-twist terms are suppressed in that their separate contributions to the perturbative curve to be compared to the large $x$ data is negligible. 
However, nonperturbative effects are present as they become absorbed in the coupling's infrared behavior.

The extracted coupling we obtain is consistent with schemes of scale fixing  
 in which  $\alpha_s$ can be extended to the entire $Q^2$ domain. Various frameworks have been proposed where the effective coupling is free of the Landau pole, {\it e.g.}, the BLM scheme \cite{Brodsky:1982gc} and its recent extension using the Principle of Maximum Comformality \cite{Wu:2013ei},   methods based on the analyticity properties of $\alpha_s$ \cite{Shirkov:1997wi,Milton:1997mi}, and finally, including nonperturbative effects  \cite{Cornwall:1981zr,Fischer:2003rp,Prosperi:2006hx,Aguilar:2009nf} (see also Refs.~\cite{Courtoy:2011mf,Courtoy:2011qa} where $\alpha_s$  was defined introducing nonperturbative effects fixed by a physical set of parameters).  While  the focus of the present manuscript is on suggesting  a way of extracting the running coupling from data, more detailed future studies will be dedicated to connecting our approach to the mentioned schemes. 

\vspace{0.5cm}
\noindent {\bf 2.} 
In order to evaluate the effect of LxR  we  perform a fit of all available large $x$, $eP$ inclusive scattering data. 
We start from standard parametrizations of the PDFs, and we consider systematically the effects of TMCs, and perturbative evolution using either NLO or next-to-leading log (NLL) resummed  coefficient functions, {\it i.e.}, with and without LxR. 
Note that the scope of the present fit is not towards a global analysis, but to assess the possible interplay among the different components that impact $Q^2$ evolution at large $x$, including LxR, TMCs, and HTs.
In the resonance region, $W^2 \leq 4$ GeV$^2$, we consider averages of both data and theoretical evaluations by comparing limited intervals 
 defined as,
\begin{eqnarray}
R^{\mbox{\tiny exp/th}}(Q^2)&=&\frac{
\int_{x_{\mbox {\tiny min}}(W^2= 4 \mbox {\scriptsize GeV}^2)}^{x_{\mbox {\tiny max}}(W^2=1.2 \mbox {\scriptsize GeV}^2)} dx\,
F_2^{\mbox {\tiny exp}} (x, Q^2)
}
{\int_{x_{\mbox{\tiny min}}(W^2=4 \mbox {\scriptsize GeV}^2)}^{x_{\mbox{\tiny max}}(W^2=1.2 \mbox {\scriptsize GeV}^2)} dx\,
F_2^{\mbox {\tiny th}} (x, Q^2)
}\quad.
\label{eq:ratio_1_1}
\end{eqnarray}

In the present analysis, we use, for $F_2^{\mbox {\tiny exp}} $, the data from JLab (Hall C, E94110)~\cite{Liang:2004tj} reanalyzed (binning in $Q^2$ and $x$) as explained in~\cite{Monaghan:2012et} as well as the SLAC data~\cite{Whitlow:1991uw}. The values of $Q^2$ and the average values of $x$ for each interval are given in Table~\ref{tab:data}. The function $F_2^{\mbox {\tiny th}}$ is the theoretical evaluation which is the same in both the DIS 
and resonance, Eq.~(\ref{eq:ratio_1_1}), regions. Notice that if Eq.~(\ref{eq:ratio_1_1}) is equal to $1$, duality is fulfilled. Since $x$ is integrated over the entire resonance region, we are considering {\em global duality}.

The OPE formulation of quark-hadron duality~\cite{Rujula:1976tz} suggests that the higher-twist contributions to the scaling structure function would either be small or cancel  otherwise duality would be strongly violated. 
However, the role of the higher-twist terms is still unclear since they would otherwise be expected to dominate the cross section at $x \rightarrow 1$.
To answer the question of the nature of a dual description, two complementary approaches have been adopted. The first is the nonperturbative model's view on the scaling of the structure functions at low-energies \cite{Close:2002tm,Jeschonnek:2003sb,Jenkovszky:2012dc}; the second approach consists in a perturbative analysis~\cite{Liuti:2011rw,Bianchi:2003hi,Liuti:2001qk}, that through LxR provides a scenario by which the effect of  HTs can be suppressed in a fully quantitative 
fit at large $x$. 
It is this second approach that we will follow in this paper.

We evaluate 
$F_2^{\mbox {\tiny th}}$ taking into account perturbative evolution at NLO, and introduce subsequently the effects of TMCs, and LxR. 
Since only valence quarks distributions are relevant in our kinematics, we consider only the Non Singlet (NS) sector,
\begin{eqnarray}
F_2^{NS} (x, Q^2) 
&=&x q(x,Q^2)+ \frac{\alpha_s}{4\pi} \mathlarger{\sum}_q \mathlarger{\int}_x^1 dz \, B_{\mbox{\tiny NS}}^q(z) \, \frac{x}{z}\, q\left(\frac{x}{z},Q^2\right)\quad,
\label{eq:convo}
\end{eqnarray}
The PDFs, $q(x, Q^2)$, are  taken from current parametrizations. We have chosen to present results using the MSTW08 set at NLO as initial parametrization~\cite{Martin:2009iq}. We have checked that there were no significant discrepancies when using other sets, {\it i.e.}, CTEQ6~\cite{Pumplin:2002vw} and the dynamical GJRFVNS~\cite{Gluck:2007ck}. The function $B_{\mbox{\tiny NS}}^q$ is the Wilson coefficient function for quark-quark.

\begin{figure}[h]
\begin{tabular}{cc}
	\includegraphics[scale= .7]{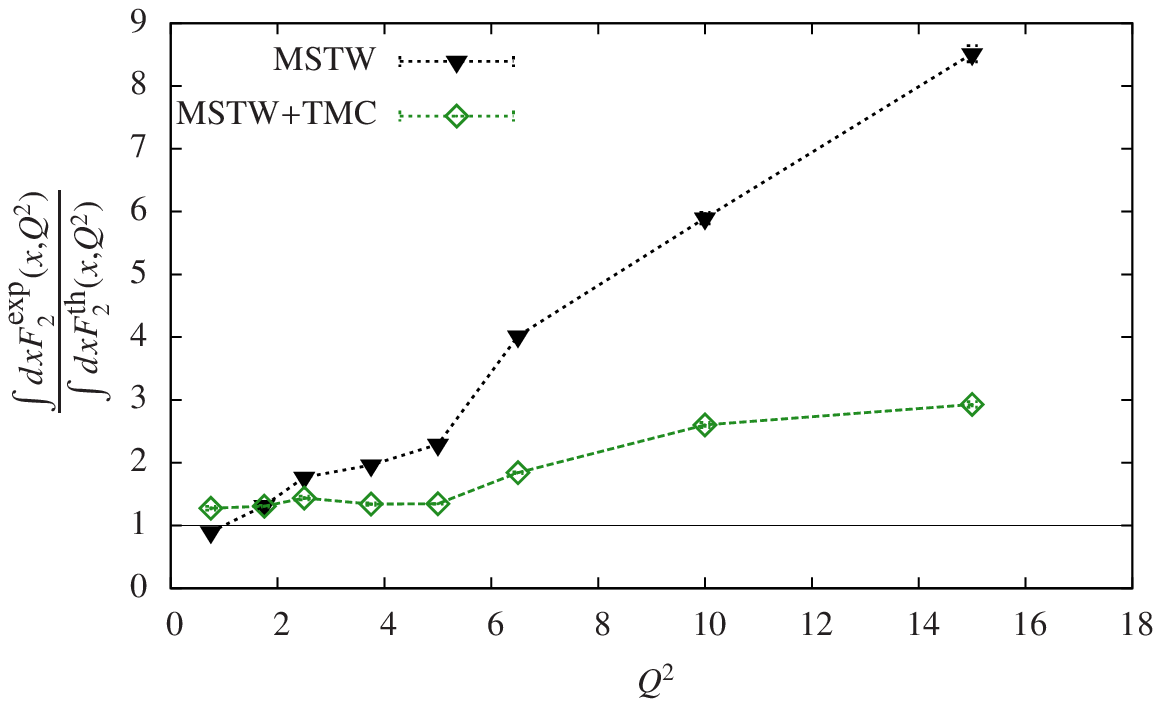}  &
	\includegraphics[scale= .7]{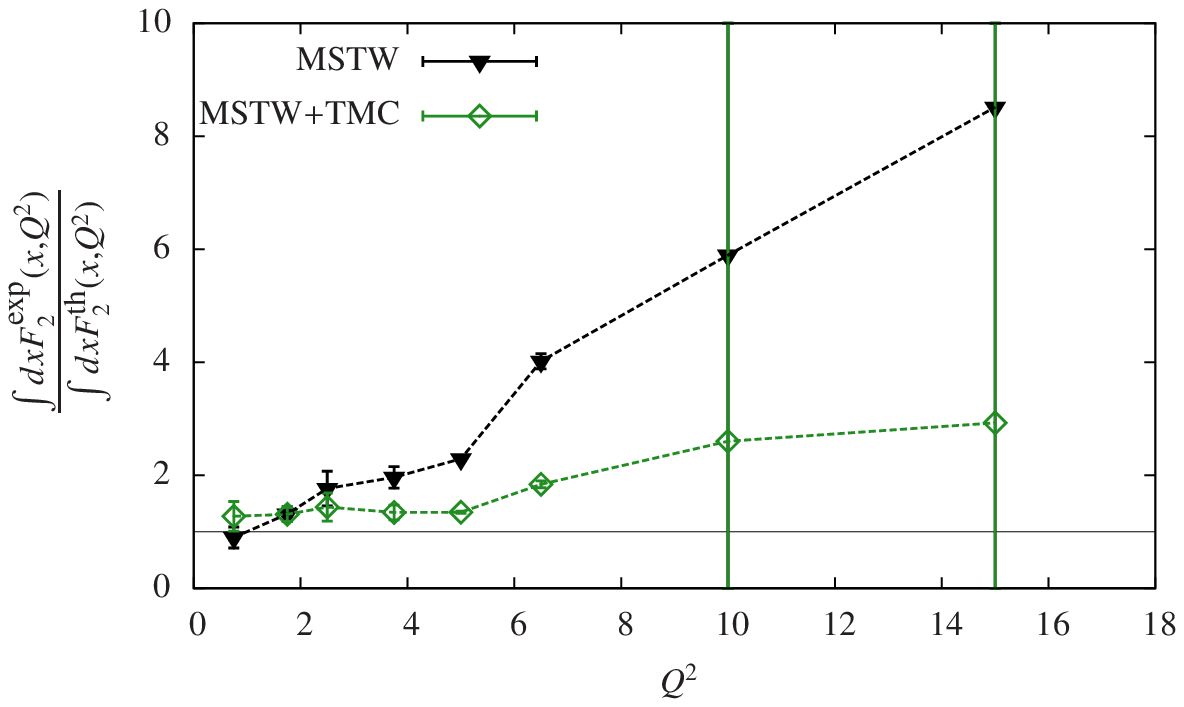}
\end{tabular}
\caption{
Ratio $R^{\mbox{\tiny exp/th}}(Q^2)$ of Eq.~(\ref{eq:ratio_1_1}) where the theoretical analysis includes PQCD evolution using the MSTW08 PDF set (black triangle), and MSTW08 PDF set plus TMCs (open green diamonds). Ratios with error bars on data integrated in quadrature ; right panel: ratios with weighted error bars on data integrated in quadrature. 
}
\label{fig:ratio}
\end{figure}

By evaluating the ratios $R{\mbox{\tiny exp/th}}$, using current parametrizations, one finds a sensible deviation from the data, even when the theoretical uncertainty from the parametrizations is included  (Figure \ref{fig:ratio}). 
One possible explanation is in the lack of accuracy in the PDF parametrizations in the large $x$, low $W^2$ domain,  since most groups implement much larger thresholds for $W^2$.
The way to a fully quantitative fit would then start from re-fitting the large $x$ data with new appropriate sets of PDFs, and simultaneously accounting for both TMCs, and LxR. The number of parameters, and the uncertainty  associated with this procedure would however be dauntingly increasing. For this reason, it is therefore necessary to take the preparatory step, conducted with the present analysis, of assessing the relative weight of the different contributions.

The additional corrections 
due to the finite mass of the initial nucleon, or the TMCs, are included directly in $F_2^{NS,\mbox {\tiny th}}$ as  \cite{Georgi:1976ve} (see also review in \cite{Schienbein:2007gr}),
\begin{eqnarray}
\label{TMC}
F_{2}^{NS (TMC)}(x,Q^2) & = &
    \frac{x^2}{\xi^2\gamma^3}F_2^{\mathrm{\infty}}(\xi,Q^2) + 
    6\frac{x^3M^2}{Q^2\gamma^4}\int_\xi^1\frac{d \xi'}{{\xi'}^2} 
F_2^{\mathrm{\infty}}(\xi',Q^2),
\end{eqnarray}
where $F_2^{\infty} \equiv F_2^{NS}$
is the structure function in the absence of TMC. Since TMC should in 
principle be applied also to the HT, we disregard 
terms of ${\cal O}(1/Q^4)$ \cite{Alekhin:2003qq}.  
Note that the expansion in Eq.~(\ref{TMC}) is valid for  $Q^2$ larger than $\approx $ 1 GeV$^2$. 
TMCs move the ratio closer to unity, as represented by the open green diamonds in Fig.~\ref{fig:ratio}. 
Uncertainties on TMCs are very small \cite{Alekhin:2003qq}. However a larger error might arise from the procedure used to account for TMCs~\cite{Accardi:2008ne}. Studies of the sensitivity to this procedure are on their way and will be published elsewhere.
At this stage, by including only TMCs and standard PDF parametrizations, we still observe  a large discrepancy with the data.

\begin{table}[h]
  \centering
  \begin{tabular}{| p{2cm} | p{2cm} | p{2cm} |}
 \hline
$Q^2$ 	[GeV$^2$]  & $x_{\mbox{\tiny ave}}$   			 &    $I^{\mbox{\tiny exp}}(Q^2)$  \\
 \hline
$1.75$\quad		&	$0.516$            &	$6.994\times 10^{-2}$	\\
$2.5$			&	$0.603$       	&	$4.881\times 10^{-2}$	\\
$3.75$			&	$0.702$        	&	$2.356\times 10^{-2}$	\\
$5.$				&	$0.753$        	&	$1.267\times 10^{-2}$	\\
$6.5$			&	$0.800$        	&	$0.685\times 10^{-2}$	\\
\hline
$4.$\quad			&	$0.712$		&	$2.045\times 10^{-2}$	\\
$5.$				&	$0.755$		&	$1.255\times 10^{-2}$	\\
$6.$				&	$0.787$		&	$0.802\times 10^{-2}$	\\
$7.$				&	$0.812$		&	$0.531\times 10^{-2}$	\\
$8.$				&	$0.832$		&	$0.363\times 10^{-2}$	\\
\hline
 \end{tabular}
 \caption{Upper block: Integrals of JLab data from Refs.~\cite{Liang:2004tj,Monaghan:2012et}, appearing in the numerator of Eq.~(\protect\ref{eq:ratio_1_1}). The first column shows the average values of $x$ for each bin. Lower block: SLAC data~\cite{Whitlow:1991uw}.}
  \label{tab:data}
\end{table}

Next, we consider LxR effects. A major consequence of LxR is a shift of the scale at which $\alpha_s$ is calculated
to lower values, with increasing $z$ (see for instance  Refs.~\cite{Brodsky:1979gy,Pennington:1982kr,Roberts:1990ww}).
This introduces a  model dependence within the PQCD approach in that  the value of the QCD running coupling  in the 
infrared region is regulated by LxR so as to satisfy duality.
In other words, LxR contains an additional degree of freedom, gathered in the definition of the coupling constant, to tune the scaling structure functions.

LxR  arises formally from terms containing powers of 
$\ln (1-z)$, $z$ being the longitudinal 
variable in the evolution equations, that are present in 
the Wilson coefficient functions $B_{\mbox{\tiny NS}}^q(z)$, in Eq.~(\ref{eq:convo}).
 To NLO and in the $\overline{\mbox{MS}}$ scheme, the Wilson coefficient function for quarks reads,
\begin{eqnarray}
B_{\mbox{\tiny NS}}^q(z)&=& \left[
\hat{P}_{qq}^{(0)}(z)\, \left\{
\ln\left(\frac{1-z}{z}\right)-\frac{3}{2}
\right\}
+\mbox{E.P.}
\right]_+\quad,
\end{eqnarray}
where E.P. means end points and $[\ldots]_+$ denotes the standard plus-prescription. The function $\hat{P}_{qq}^{(0)}(z)$ is the LO splitting function for quark-quark.
The logarithmic terms, {\it i.e.}, $\ln(1-z)$,  in $B_{\mbox{\tiny NS}}^q(z)$ become very large at large $x$ values. They need to be 
resummed to all orders in $\alpha_s$. 
Resummation was first introduced by  
linking this issue to the definition of the correct kinematical variable that determines the 
phase space for the radiation of gluons
at large $x$. This was found to be $\widetilde{W}^2 = Q^2(1-z)/z$, 
instead of $Q^2$~\cite{Brodsky:1979gy,Amati:1980ch}.
As a result, the argument of the strong coupling constant becomes $z$-dependent~\cite{Roberts:1999gb},
\begin{equation} 
\alpha_s(Q^2) \rightarrow \alpha_s\left(Q^2 \frac{(1-z)}{z}\right)\quad.
\end{equation}
In this procedure, however, an ambiguity is introduced, related to the need of continuing 
the value of $\alpha_s$  
for low values of its argument, {\it i.e.}, for $z \rightarrow 1$~\cite{Pennington:1981cw}. 
Since the size of this ambiguity is of the same order as the higher-twist corrections, it has been considered, in a previous work~\cite{Niculescu:1999mw}, as a source of theoretical error or higher order effects. 
We investigate the effect induced by changing the argument of $\alpha_s$ on the behavior of the $\ln(1-z)$-terms in the convolution Eq.~(\ref{eq:convo}), and  resum those terms as
\begin{eqnarray}
\ln(1-z)&=&\frac{1}{\alpha_{s, \mbox{\tiny LO}}(Q^2) }\int^{Q^2} d\ln Q^2\, \left[\alpha_{s, \mbox{\tiny LO}}(Q^2 (1-z)) -\alpha_{s, \mbox{\tiny LO}}(Q^2) \right]
\equiv \ln_{\mbox{\tiny LxR}}\quad,
\end{eqnarray}
 including the complete $z$ dependence of $\alpha_{s, \mbox{\tiny LO}}(\tilde W^2)$ to all logarithms.\footnote{The terms proportional to $\ln z$ are not divergent at $z\to 1$.}
Note that we are using three different concepts of order expansions. The present analysis is conducted to \emph {next-to-leading order} (we evolve the PDF sets to NLO), to \emph {leading-twist} (we consider the LT PDFs only) and to \emph {all logarithms} (we include $\alpha_{s, \mbox{\tiny LO}}($scale$)$ to all logarithms).
 This resummation is easily understood when considering the first term of the expansion of $\alpha_s(\tilde W^2)$ in $\ln((1-z)/z)$,
	\begin{eqnarray}
		\alpha_s(\tilde W^2)=\alpha_s(Q^2)-\frac{\beta_0}{4\pi}\, \ln\left(\frac{1-z}{z}\right)\, \alpha_s^2(Q^2),
	\end{eqnarray}	
as proposed in Ref.~\cite{Roberts:1999gb}.
To all logarithms, the convolution becomes
\begin{equation}
F_2^{NS,  \mbox{\tiny Resum}} (x, Q^2)  =x q(x,Q^2)+ \frac{\alpha_s}{4\pi} \mathlarger{\sum}_q \mathlarger{\int}_x^1 dz \, B_{\mbox{\tiny NS}}^{ \mbox{\tiny Resum}}(z) \, \frac{x}{z}\, q\left(\frac{x}{z},Q^2\right),
\label{eq:convo_eff}
\end{equation}   
where,
\begin{equation}
B_{\mbox{\tiny NS}}^{ \mbox{\tiny Resum}} =  B_{\mbox{\tiny NS}}^q(z) - \hat{P}_{qq}^{(0)}(z)\, \ln(1-z)+\hat{P}_{qq}^{(0)}(z)\,\ln_{\mbox{\tiny LxR}}.
\end{equation}

Using $F_2^{NS,  \mbox{\tiny Resum}}$ plus TMCs, in Eq.~(\ref{eq:ratio_1_1}), will make the ratio $R$ decreases substantially, essentially  leaving no space for HT terms. 
This is due in our approach mostly to the change of the argument of the running coupling constant. 
At fixed $Q^2$, in the integration over $x<z<1$, the scale $\widetilde{W}^2= Q^2 (1-z)/z$ is shifted  and can reach low values, where the running of the coupling constant starts blowing up. At this stage, our analysis requires nonperturbative information.
A way to address this issue is to set a maximum value for the longitudinal momentum fraction, $z_{\mbox{\tiny max}}$, which defines a limit from which nonperturbative effects have to be accounted for, and to cut $\alpha_s$ at the corresponding scale, $\widetilde{W}^2 (z_{\mbox{\tiny max}})= Q^2 (1-z_{\mbox{\tiny max}})/z_{\mbox{\tiny max}}$. Larger values of $z_{\mbox{\tiny max}}$ correspond to lower values at which the scale should be cut in the analysis, meaning that the perturbative value can be used. As we show later, large $z_{\mbox{\tiny max}}$ occurs in the data at large $Q^2$, therefore the effect of the shift in scale gets smaller.  

The functional form $\ln_{\mbox{\tiny LxR}}$ is therefore slightly changed. Two distinct regions can be studied: the ``running" behavior  in $x<z<z_{\mbox{\tiny max}}$ and the ``steady" behavior $z_{\mbox{\tiny max}}<z<1$, 
\begin{eqnarray}
F_2^{NS,  \mbox{\tiny Resum}} (x, z_{\mbox{\tiny max}}, Q^2)  &=&x q(x,Q^2)+ \frac{\alpha_s}{4\pi} \mathlarger{\sum}_q 
\left \{
\mathlarger{\int}_x^1 dz \, \left[B_{\mbox{\tiny NS}}^q(z) - \hat{P}_{qq}^{(0)}(z)\, \ln(1-z)\right]\, \right.\nonumber\\
&&\left.+  \mathlarger{\int}_x^{z_{\mbox{\tiny max}}} dz \, \hat{P}_{qq}^{(0)}(z)\,\ln_{\mbox{\tiny LxR}}+ \ln_{\mbox{\tiny LxR, max}}\, \mathlarger{\int}_{z_{\mbox{\tiny max}}}^1 dz \, \hat{P}_{qq}^{(0)}(z)\,
\right\} \frac{x}{z}\, q\left(\frac{x}{z},Q^2\right). \nonumber \\
\quad 
\label{eq:convo_max}
\end{eqnarray}   
$z_{\mbox{\tiny max}}$ appears therefore as a free parameter in our analysis. A possible criterion to constrain it is to fit the large $x$ data assuming a null direct contribution to the structure function from the dynamical HTs namely, for each $Q^2$ bin we define $z_{\mbox{\tiny max}}$ by varying $R^{\mbox{\tiny exp/th}}$ as a function of $z_{\mbox{\tiny max}}$,  so that
\begin{eqnarray}
R^{\mbox{\tiny exp/th}}(z_{\mbox{\tiny max}}, Q^2)
&=&\frac{
 \mathlarger{\int}_{x_{\mbox {\tiny min}}}^{x_{\mbox {\tiny max}}} dx\,
F_2^{\mbox {\tiny exp}} (x, Q^2)
}
{ \mathlarger{\int}_{x_{\mbox{\tiny min}}}^{x_{\mbox{\tiny max}}} dx\,
F_2^{NS, \mbox{\tiny Resum}} (x, z_{\mbox{\tiny max}}, Q^2)
}
= \frac{I^{\mbox{\tiny exp}}}{I^{ \mbox{\tiny Resum}}}= 1\quad.
\label{eq:ratio_max}
\end{eqnarray}
In Eq.~(\ref{eq:ratio_max}), $F_2^{NS,  \mbox{\tiny Resum}} (x, z_{\mbox{\tiny max}}, Q^2)$ was evaluated including TMCs, resummation to all log, and setting possible
dynamical HT contributions to zero. 
The latter get, however, absorbed in the coupling's infrared behavior.
More precisely, the suppression of HTs in the structure function is compensated by 
 the behavior of $\alpha_s$ in the infrared region.
As a result, contrarily to what originally deduced in, 
{\it e.g.}, Ref.~\cite{Niculescu:2000tk}, a definite role of nonperturbative corrections 
is obtained, pointing at the fact that duality, defined on the basis of a
dominance of single parton scattering, {\it i.e.}, suppression of final state interactions, might indeed be broken. 

\begin{figure}[h]
\begin{tabular}{cc}
	\includegraphics[scale= .7]{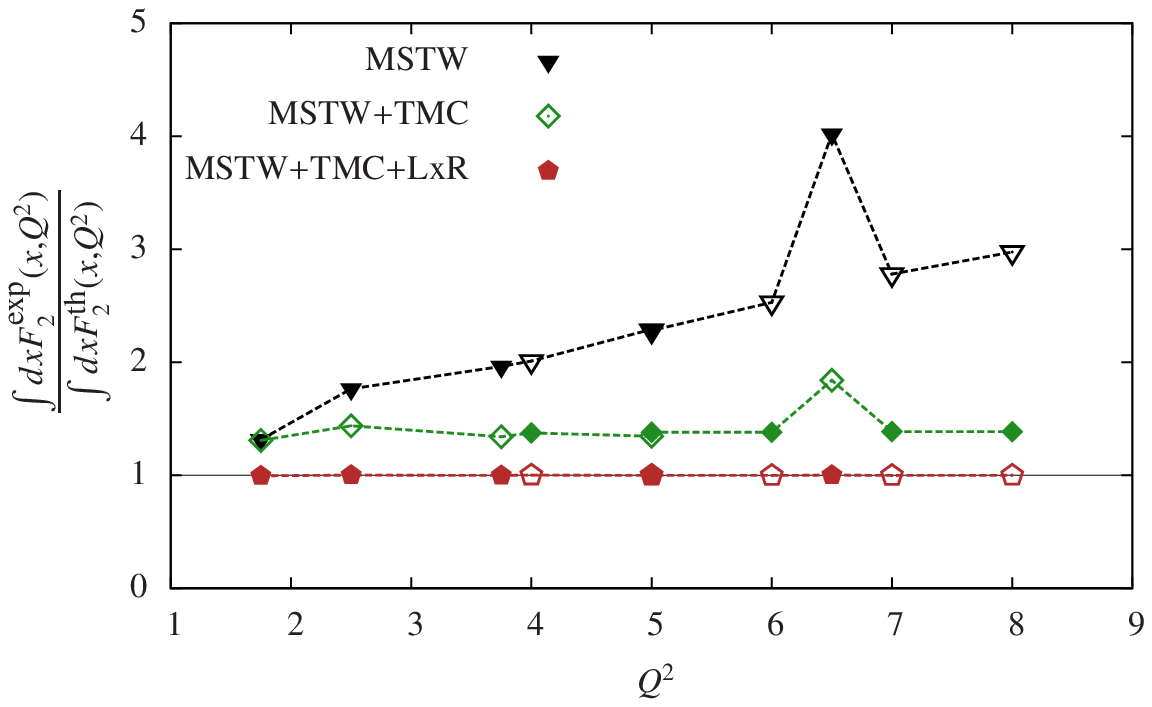}&
	\includegraphics[scale= .7]{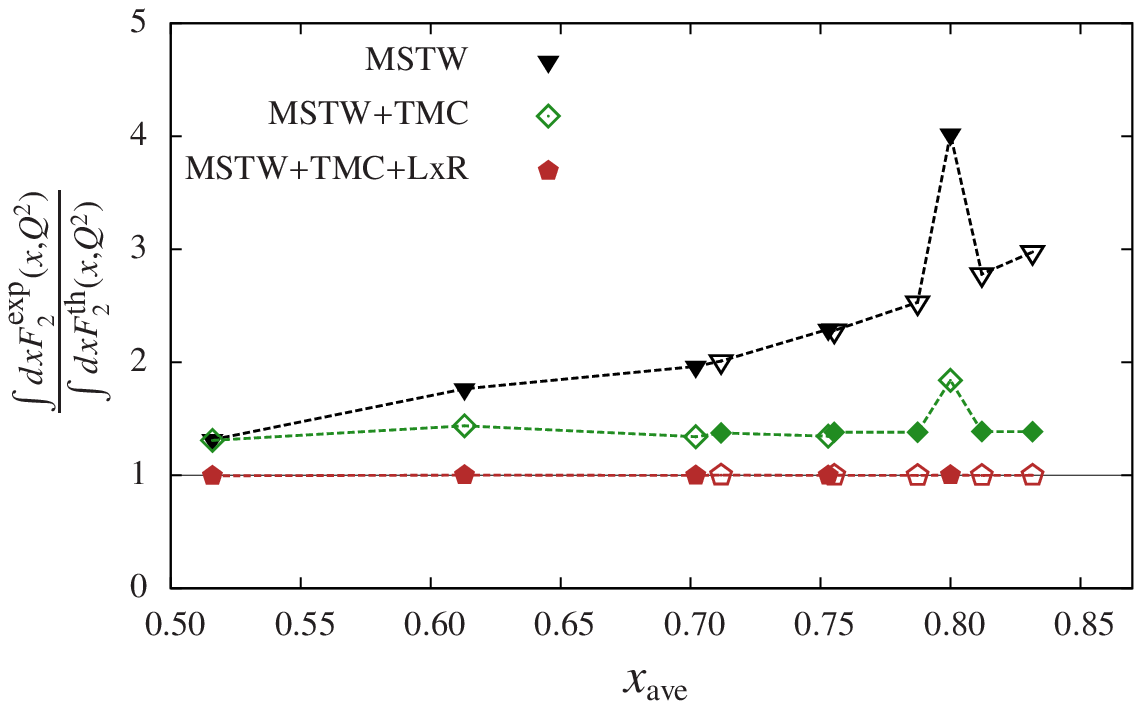}
\end{tabular}
\caption{The ratio $R^{\mbox{\tiny exp/th}}(x_{\mbox{\tiny ave}}, Q^2)$ as a function of $Q^2$, on the left pannel, and as a function of $x_{\mbox{\tiny ave}}$ on the right. Same as Fig.~\ref{fig:ratio} but with the red hexagon representing the LxR results of Tab.~\ref{tab:lxr}. The key shows the results corresponding to JLab data. The open triangle, full diamonds and open hexagons corresponds to SLAC data.}
\label{fig:ratio_lxr}
\end{figure}

Results are represented by the red hexagons in Fig.~\ref{fig:ratio_lxr}. The integrals values  are given in Tab.~\ref{tab:lxr} together with the corresponding values for $z_{\mbox{\tiny max}}$.
Since, for the largest values of $Q^2$,  $Q^2=10, \,15$GeV$^2$, outside the resonance region, on Fig.~\ref{fig:ratio}, $z_{\mbox{\tiny max}}$ becomes closer to 1, we do not consider those data points in what follows.


\begin{table}
  \centering
  \begin{tabular}{| p{2cm} | p{2.5cm} | p{2.5cm} | p{2.7cm} | p{3.cm} | p{1.5cm} |}
 \hline

    $Q^2$ 	[GeV$^2$]  & $I^{\mbox{\tiny exp}}(Q^2)$  &   $I^{(0),\mbox{\tiny DIS}}(Q^2)$	&   $I^{(0),\mbox{\tiny DIS+TMC}}(Q^2)$     &  $I^{ \mbox{\tiny Resum}}(z_{\mbox{\tiny max}}, Q^2)$  & $z_{\mbox{\tiny max}}$\\
 \hline
	$1.75$     &	$6.994\times 10^{-2}$	&	$5.316\times 10^{-2}$     &	$5.345\times 10^{-2}$	          &	$7.025\times 10^{-2}$		&   	  $0.63$\\
	$2.5$	&	$4.881\times 10^{-2}$	&	$2.765\times 10^{-2}$     &	$3.393\times 10^{-2}$           	  &	$4.872\times 10^{-2}$		&   	  $0.745$\\
	$3.75$     &	$2.356\times 10^{-2}$	&	$1.201\times 10^{-2}$     &	$1.756\times 10^{-2}$	          &	$2.359\times 10^{-2}$		&   	  $0.76$\\
	$5.$		&	$1.267\times 10^{-2}$	&	$0.553\times 10^{-2}$     &	$0.942\times 10^{-2}$	          &	$1.270\times 10^{-2}$		&   	  $0.79$\\
	$6.5$	&	$0.685\times 10^{-2}$	&	$0.170\times 10^{-2}$	&	$0.372\times 10^{-2}$           	   &	$0.683\times 10^{-2}$   		&	 $0.9$	\\
\hline
	$4.$		&	$2.045\times 10^{-2}$	&	$1.017\times 10^{-2}$	&	$1.487\times 10^{-2}$           	   &	$2.041\times 10^{-2}$   		&	 $0.79$	\\
	$5.$		&	$1.255\times 10^{-2}$	&	$0.550\times 10^{-2}$	&	$0.909\times 10^{-2}$           	   &	$1.255\times 10^{-2}$   		&	 $0.811$	\\
	$6.$		&	$0.802\times 10^{-2}$	&	$0.317\times 10^{-2}$	&	$0.581\times 10^{-2}$           	   &	$0.803\times 10^{-2}$   		&	 $0.825$	\\
	$7.$		&	$0.531\times 10^{-2}$	&	$0.191\times 10^{-2}$	&	$0.383\times 10^{-2}$           	   &	$0.532\times 10^{-2}$   		&	 $0.837$	\\
	$8.$		&	$0.363\times 10^{-2}$	&	$0.122\times 10^{-2}$	&	$0.262\times 10^{-2}$           	   &	$0.363\times 10^{-2}$   		&	 $0.845$	\\
  \hline
 \end{tabular}
 \caption{Integrals at each stage. In the last columns: the value $z_{max}$ associated with $I^{ \mbox{\tiny Resum}}(z_{\mbox{\tiny max}}, Q^2)$.}
  \label{tab:lxr}
\end{table}

\vspace{0.5cm}
\noindent {\bf 3.}
Based on the results of our analysis of large $x$ data including TMCs and LxR, 
we now extract $\alpha_s$  by assuming that it runs from the onset of a minimal scale which is determined from the comparison with data,
and it is frozen from that minimal scale downward to the real photon limit (scale=0 GeV$^2$).
As one can see from Table~\ref{tab:lxr},  data in the resonance region are crucial for this determination. 

\begin{figure}[h]
\includegraphics[scale= 1.]{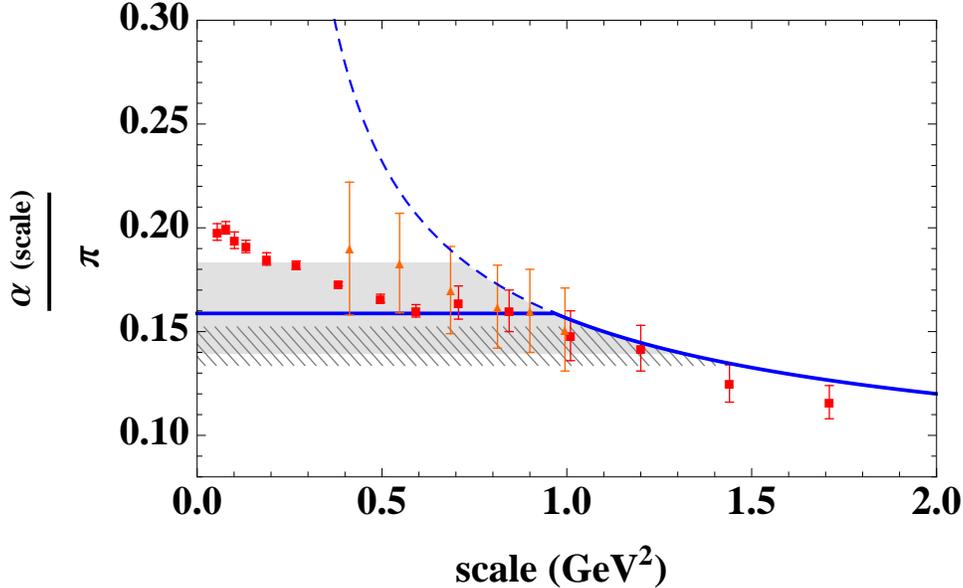}
\caption{(color online) Extraction of $\alpha_s$. The blue dashed curve represents the exact NLO solution for the running coupling constant in $\overline{\mbox{MS}}$ scheme. The solid blue curve represents the coupling constant obtained from our analysis using inclusive electron scattering data at large $x$.  Owing to large $x$ resummation, at  lower values of the scale, $\alpha_s=\alpha_{s, \mbox{\tiny NLO}}\left(\mbox{min} \right)$ is frozen as explained in the text. The grey area represents the region where the freezing occurs for JLab data, while the hatched area corresponds the freezing region determined from SLAC data. This  error band represents  the theoretical uncertainty in our analysis. We also plot results extrapolated \cite{alexandre} from the recent analysis in Refs.~\cite{Deur:2005cf, Deur:2008rf}: the red squares correspond to $\alpha_s$ extracted from Hall B CLAS EG1b, with statistical uncertainties; the  orange triangles corresponds to Hall A E94010 / CLAS EG1a
data, the uncertainty here contains both statistics and systematics.}
\label{extract}
\end{figure}

In  Fig.~\ref{extract} we show our extracted value $\alpha_{s, \mbox{\tiny NLO}}$(scale) where we used the $\overline{\mbox{MS}}$ scheme outside the IR region, for the same value of $\Lambda$ throughout this paper. $\alpha_s$ was obtained as an exact solution to NLO \cite{Courtoy:2011mf}.  
Our theoretical error band is defined by the shift in $z_{\mbox{\tiny max}}$ from the different bins displayed in Table~\ref{tab:lxr} namely, 
\begin{equation}
\alpha_{s, \mbox{\tiny NLO}}\left(Q_i^2\frac{(1-z_{\mbox{\tiny max}, i})}{z_{\mbox{\tiny max}, i }}\right) \qquad \mbox{for}\qquad i=1,\ldots 10 \qquad,
\end{equation}
$i$ corresponds to the data points.
Including this error band, our extracted frozen value of the coupling constant is, using the MSTW08 PDF set for the analysis,
\begin{equation}
0.1337\leq \frac{\alpha_{s, \mbox{\tiny NLO}} (\mbox{scale}\to 0 \mbox{GeV}^2)}{\pi} \leq 0.1839 \quad.
\end{equation}
%
%
In the figure we also report  values from the extraction using polarized $eP$ scattering data in Ref.~\cite{Deur:2005cf, Deur:2008rf,Brodsky:2010ur,alexandre}. These values represent the  first extraction of an effective coupling in the IR region that was obtained by analyzing the data  
relevant for the study of the GDH sum rule. To extract the coupling constant,  the $\overline{\mbox{MS}}$ expression of the Bjorken sum rule up to the 5th order in alpha (calculated in the $\overline{\mbox{MS}}$ scheme) was used. In order to compare with our extraction using the $F_2^p$ observable, the finite value for $\alpha_s(0)$  found in \cite{Deur:2005cf, Deur:2008rf,Brodsky:2010ur} was rescaled in \cite{alexandre} assuming the validity of the commensurate scale relations \cite{Brodsky:2010ur} in the entire range of the scale entering the analysis. 
The agreement with our analysis which is totally independent, is impressive.


\vspace{0.5cm}
\noindent {\bf 4.}
 In conclusion, we presented an extraction of $\alpha_s$ using $eP$ scattering data at large $x$. A careful analysis of all the contributions appearing at large $x$ including TMCs and LxR, was performed. The central value for $\alpha_s(Q^2<1$GeV$^2)/\pi$ was found to be $0.1588$. This value is in agreement with the extraction from the GDH sum rule analysis ~\cite{Deur:2005cf, Deur:2008rf, alexandre}.

When considering PQCD observables at low scales, we implicitly face an interpretation problem. In the multi-GeV$^2$ region and at large-$x$, where the resonances lie, perturbative QCD is pushed to its limits. Both higher terms in the perturbative expansion of that observable, and power corrections  need be taken into account.  In the present approach, this transition is  taken into account  by re-interpreting the running coupling constant, at the scale of transition instead.
By tuning the scaling structure functions to the averaged data in the resonance region, we parametrize the realization of duality through an infrared fixed-point below which the strong coupling constant stops running. The value we have found for the transition scale is situated around $1$ GeV$^2$. The uncertainties on that value are, however, still important. 

We report an interesting observation that by connecting the values of the coupling  using two different observables and schemes, namely the $g_1$ scheme from the GDH sum rule extraction, and our $\overline{\mbox{MS}}$ based extraction, we obtain   
an excellent agreement by simply extending the commensurate scale relations to the IR region (see also \cite{Grunberg:1980ja,Grunberg:1982fw}).
While our conclusion ensues from a perturbative analysis, in the near future, we will explore the role of nonperturbative 
effective couplings as well.
An interesting avenue is provided in this direction by the evaluation of the strong coupling constant in the infrared region using the AdS/CFT conjecture that was recently performed in Ref.~\cite{Brodsky:2010ur}.

The importance of finite couplings has been highlighted many times in the Literature, see, {\it e.g.},~\cite{Wu:2013ei}  and references therein.
Our analysis, would allow in principle, to extract from a fit the nonperturbative parameters often present in the proposed functional forms for the running of $\alpha_s$, {\it e.g.}, in Refs.~\cite{Cornwall:1981zr, Fischer:2003rp, Shirkov:1997wi}.

\vspace{1.2cm}

We are grateful to Vicente Vento and Stan Brodsky for many discussions and invaluable advice.  We also thank Alexandre Deur and Jian Ping Chen for useful discussions and for sending us  their extracted values of the coupling constant, Peter Monaghan and Donal Day for directing us to the large $x$ database. The support of Gruppo III of INFN, Laboratori Nazionali di Frascati, where part of the manuscript was completed is also wholeheartedly acknowledged. This work was funded by the Belgian Fund F.R.S.-FNRS via the contract of ChargŽe de recherches (A.C.), and by U.S. D.O.E. grant  DE-FG02-01ER4120 (S.L.).

%
\bibliographystyle{apsrevM}
\bibliography{lxr}
\end{document}